\documentclass[preprint,amsmath,amssymb,superscriptaddress,12pt,prb]{revtex4}
\usepackage{amsmath}
\usepackage{graphicx}
\usepackage{dcolumn}
\usepackage{epstopdf}
\usepackage{extarrows}
\usepackage{multirow}
\usepackage{tabularx}
\usepackage[normalem]{ulem} 

\begin{document}

\title{Metallic, Magnetic and Molecular Nanocontacts} 

\author{Ryan Requist}
\email{rrequist@mpi-halle.mpg.de}
\affiliation{International School for Advanced Studies (SISSA), Via Bonomea 265, Trieste 34136, Italy}
\affiliation{Max Planck Institute of Microstructure Physics, Weinberg 2, 06114 Halle, Germany}
\author{Pier Paolo Baruselli}
\affiliation{International School for Advanced Studies (SISSA), Via Bonomea 265, Trieste 34136, Italy}
\affiliation{Institut f\"ur Theoretische Physik, Technische Universit\"at Dresden, 01062 Dresden, Germany}
\affiliation{Democritos Simulation Center, Istituto Officina dei Materiali, Consiglio Nazionale delle Ricerche, Via Bonomea 265, Trieste 34136, Italy}
\author{Alexander Smogunov}
\affiliation{Service de Physique de l'Etat Condens\'e (SPEC), CEA, CNRS, Universit\'e Paris-Saclay, CEA Saclay 91191 Gif-sur-Yvette Cedex, France}
\author{Michele Fabrizio}
\affiliation{International School for Advanced Studies (SISSA), Via Bonomea 265, Trieste 34136, Italy}
\affiliation{Democritos Simulation Center, Istituto Officina dei Materiali, Consiglio Nazionale delle Ricerche, Via Bonomea 265, Trieste 34136, Italy}
\author{Silvio Modesti}
\affiliation{Physics Department, University of Trieste, Via Valerio 2, Trieste 34127, Italy}
\affiliation{TASC Laboratory, Istituto Officina dei Materiali, Consiglio Nazionale delle Ricerche, s.s. 14 km 163.5. Trieste 34149, Italy}
\author{Erio Tosatti}
\affiliation{International School for Advanced Studies (SISSA), Via Bonomea 265, Trieste 34136, Italy}
\affiliation{Democritos Simulation Center, Istituto Officina dei Materiali, Consiglio Nazionale delle Ricerche, Via Bonomea 265, Trieste 34136, Italy}
\affiliation{International Centre for Theoretical Physics (ICTP), Strada Costiera 11, Trieste 34151, Italy}

\date{\today}

\begin{abstract}
Scanning tunnelling microscopy and break-junction experiments realize metallic and molecular nanocontacts that act as ideal one-dimensional channels between macroscopic electrodes. Emergent nanoscale phenomena typical of these systems encompass structural, mechanical, electronic, transport, and magnetic properties. This Review focuses on the theoretical explanation of some of these properties obtained with the help of first-principles methods. By tracing parallel theoretical and experimental developments from the discovery of nanowire formation and conductance quantization in gold nanowires to recent observations of emergent magnetism and Kondo correlations, we exemplify the main concepts and ingredients needed to bring together \textit{ab initio} calculations and physical observations. It can be anticipated that diode, sensor, spin-valve and spin-filter functionalities relevant for spintronics and molecular electronics applications will benefit from the physical understanding thus obtained.
\end{abstract}

\maketitle

\begin{text}

A nanocontact is a local point of contact between two macroscopic conductors where electrical current is constrained to pass through a cross-section about a nanometre wide.  The conductance through a nanocontact is characteristically determined by the scattering on the local atomic and electronic structure of this region.  Nanocontacts may consist of a single molecule, a constriction in a metallic wire, or even a monatomic chain of atoms, down to a single atom,\cite{agrait2003} and have been fabricated in break junctions,\cite{muller1992} scanning tunnelling microscopes (STMs),\cite{gimzewski1987} transmission electron microscopes\cite{ohnishi1998,rodrigues2001} and on insulating substrates by electromigration\cite{park1999} and electrodeposition.\cite{li1999,morpurgo1999}  With modern techniques it is possible to form molecular nanocontacts which are stable for several hours and can be mechanically controlled within a break junction or by an STM tip and electrically modulated by gate and bias voltages or the immediate nanoscale environment.

Gold nanocontacts are known for the remarkable phenomenon of conductance quantization,\cite{agrait1993,pascual1993,brandbyge1995,gai1996,ohnishi1998,yanson1998} 
which also occurs in some other metals\cite{krans1993,olesen1994} and is one of the few quantum effects visible at room temperature.  The famous ductility of gold extends to the nanoscale, where it readily forms stable monatomic wires with conductance quantized in units of $2e^2/h$, where $e$ is the elementary charge and $h$ is Planck's constant.  The conductance of transition metals is more complicated, but like that of gold can still be understood within the Landauer-B\"uttiker theory of ballistic conductance.\cite{landauer1957,buettiker1988,imry1999} Spontaneous magnetism, such as that predicted\cite{delin2004b, smogunov2008a, smogunov2008b} and observed \cite{strigl2015} in platinum nanocontacts, is a prime example of ``small is different,''\cite{landman1992} that is, how nanoscale systems in reduced dimensions can behave differently than expected from what we know about their bulk material properties.  With growing interest in nanocontacts from the fields of molecular electronics\cite{aviram1974,joachim2000,tao2006,heath2009,scott2010} and spintronics,\cite{zutic2004,sanvito2011} investigations have recently broadened to molecular nanocontacts, which present additional electronic, spin and vibrational complexities, as well as many-body phenomena, such as Coulomb blockade and the Kondo effect.  

The purpose of this focused Review, with no claims of completeness, is to summarize an arc of interrelated developments stretching from foundational work on gold nanowires and conductance quantization to magnetic nanocontacts and finally dynamical magnetic correlations in the form of Kondo conductance anomalies.  Emphasis is placed on the intersection of state-of-the-art first principles electronic structure calculations and experiment in the decade following a review of metallic point contacts.\cite{agrait2003}  In view of the recent experimental verification of several theoretical predictions, it is now an ideal time to summarize anew the status of the field.  The topics covered include the structure and mechanics of metallic nanocontacts, \textit{ab initio}-based ballistic electron transport as described by the Landauer-B\"uttiker scattering formalism, magnetic nanocontacts with separate spin-up and spin-down conductance channels, noise measurements for identifying spontaneous emergent magnetism, Kondo conductance anomalies in transport through transition metal adatoms and molecules, and exotic Kondo physics involving a breakdown of the Fermi liquid picture.  Magnetic and molecular nanocontacts are of particular interest for their potential device applications as discussed in the final section.
\medskip

% 2
\noindent \textbf{Mechanics and structural properties} 

Accidental nanocontacts routinely form and break in the context of friction between bodies. Specifically planned nanocontacts may be fabricated by a variety of techniques.  In break-junction experiments, it was found that a notched wire under tension develops a narrow neck before eventually breaking.  Bringing the ends back together forms a point contact, sometimes with the cross-section of a single atom.  In STMs, a sharp metal tip is pushed into a metallic surface and gradually retracted until the contact has shrunk to a single atom.  Just before the breaking point, the conductance of these nanocontact systems decreases in approximately quantized steps. The closeness of the very last steps to multiples of $2e^2/h$ indicates that the last contact is a single atom wide.  All this early work has been superbly reviewed by Agra\"{i}t and colleagues.\cite{agrait2003}  Experimental work has been paralleled by numerical simulations,\cite{landman1990,sutton1990,guelseren1998,nakamura1999, hakkinen2000, bahn2001,dasilva2004,coura2004,dednam2015} usually of the molecular dynamics type, where interatomic forces are provided by one or another force field.  Since simulations can only handle extremely short times, it is important to address the general issue of timescales.

Conceptually, the process of breaking a ductile metallic nanocontact under traction involves a multiplicity of timescales.  The first and shortest, $\tau_1$, is that of initial plastic yield, slippage, atom rearrangement, and a very system-specific local thinning with destruction of the bulk crystalline structure. On a second, longer timescale, $\tau_2$, a slower flow of atoms may, if ductility is sufficient, establish a nearly steady state between the bulk-like ``leads" and the nanocontact region.  As we shall see below, in ductile metals such as gold, platinum, and some alkalis, the nanocontact may adopt well-defined, long-lived even if temporary, near-equilibrium geometrical shapes, such as nanowires.  The nanocontact is in this case under an effective tension (even in the absence of traction) which drives an associated flow of atoms away from the contact and into the bulk-like leads.  During this spontaneous thinning process, the nanocontact evolves through thinner and thinner shapes, lingering for longer times in those whose ``magic" structures are energetically favourable.  Alternatively, in less interesting non-ductile materials, there is no time to attain this metastable stage before the third timescale, $\tau_3$, is reached, where the nanocontact snaps and breaks.  Time scale $\tau_1$ may be realistically reached and atomistically described in real time by molecular dynamics simulations. The longer timescale $\tau_2$ is instead generally beyond reach. 

As an alternative to simulations, the process of adiabatic nanocontact thinning can be given a thermodynamic description, which is revealing even if approximate.  Let us discuss here the case of nanowires.  A phenomenon that attracted attention is the spontaneous appearance during pulling, particularly in gold, of metallic wires with quantized diameters and shapes, reaching a length of several nanometres before breaking.  Physical reasons behind the formation of nanowires were given in Ref.~\onlinecite{tosatti2005}, following earlier proposals,\cite{torres1999,tosatti2001} as well as in Ref.~\onlinecite{hakkinen2000}.  

Simulations had predicted\cite{guelseren1998} that free nanowires could adopt ``weird'' non-crystalline helical shapes below a critical radius.  A spectacular variety of magic (meaning reproducibly well-defined) weird nanowire shapes were discovered by Takayanagi's group in their transmission electron microscopy study of gold nanocontacts suspended between leads\cite{kondo2000} (see Fig.~\ref{fig:magic}a-d).
\begin{figure}[tb]
\begin{minipage}{\textwidth}
  \centering
  \hspace*{-2.3cm}
  \raisebox{-0.2\height}{\includegraphics[height=8.5cm]{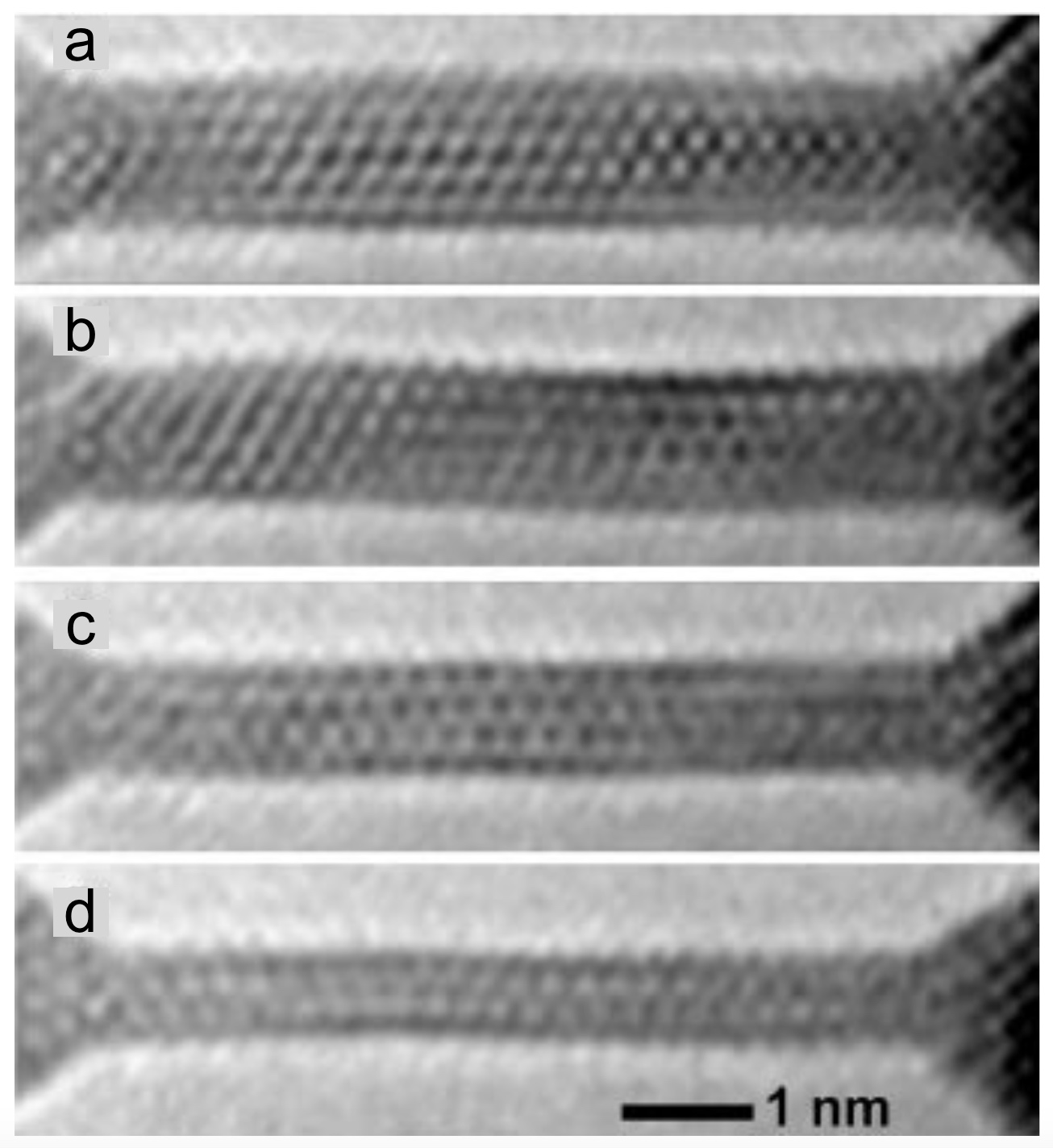}}
  \hspace*{0.0cm}
  \raisebox{0.6\height}{
  \begin{tabularx}{0.34\textwidth}{cc}
 \includegraphics[width=0.28\textwidth]{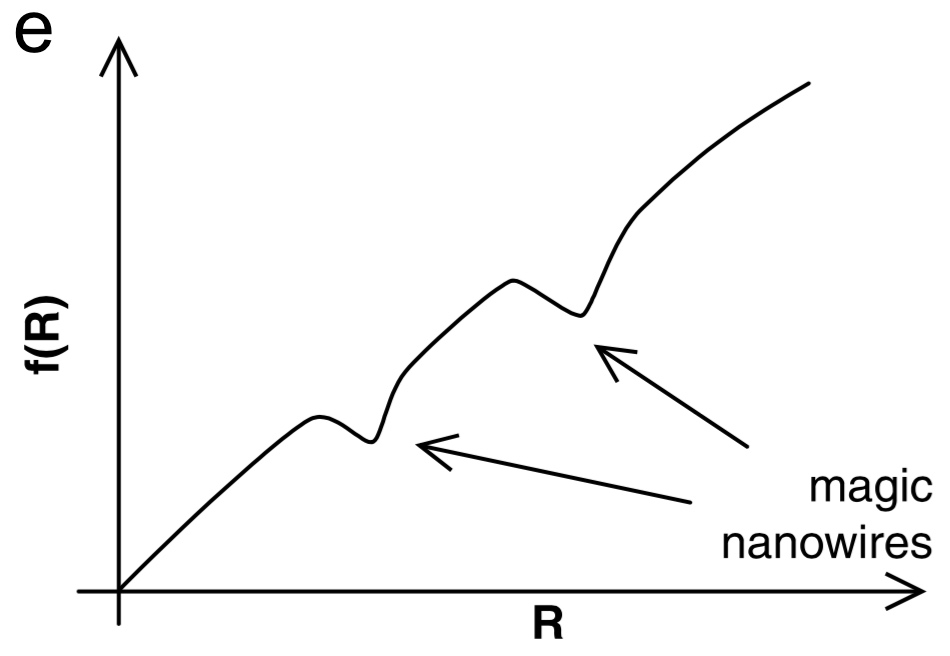} &
 \includegraphics[width=0.18\textwidth]{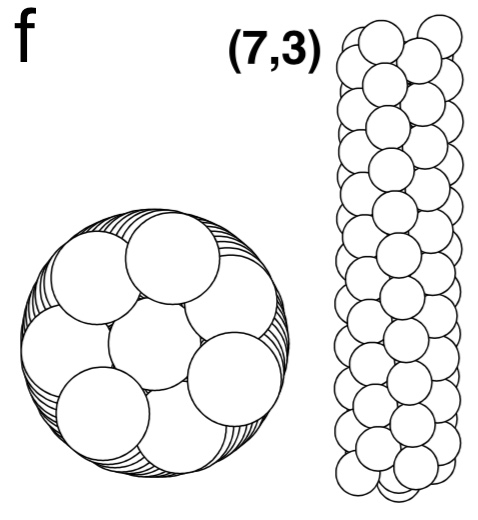} \\
 \includegraphics[width=0.18\textwidth]{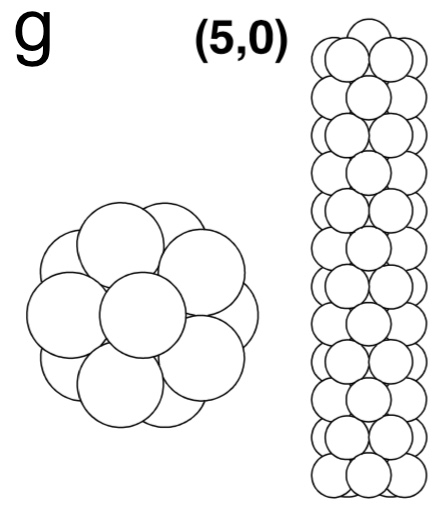} &
 \includegraphics[width=0.18\textwidth]{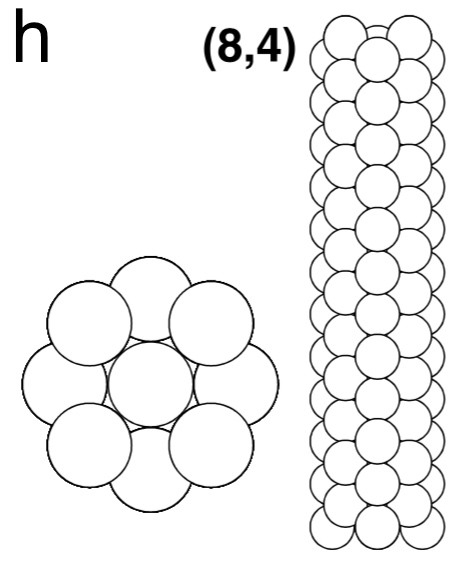}
 \end{tabularx}
  }
\end{minipage}
\caption{\textbf{Gold nanowires. a-d}, Transmission electron microscope images of progressively thinning gold nanowires with diameters of 1.3, 1.1, 0.8 and 0.6~nm, respectively. \textbf{e}, String tension $f$ of a tip-suspended nanowire as a function of radius $R$.  Globally, $f$ decreases with decreasing radius; ``magic'' structures correspond to minima of energy $E(N)$ (where $N$ is the number of atoms in the wire), reflected in local minima of $f$. \textbf{f}, The stable magic nanowire (7,3) is chiral.  \textbf{g,h}, Nanowires (5,0) and (8,4) are linear and achiral.  The notation $(m,n)$ denotes the unique nanowire tube for which $m\mathbf{a}_1+n\mathbf{a}_2$ is the unit cell vector orthogonal to the tube axis; $\mathbf{a}_1$ and $\mathbf{a}_2$ are the primitive lattice vectors of a (111) Au sheet.  Figure reproduced with permission from: \textbf{a-d}, Ref.~\onlinecite{kondo2000}, AAAS; \textbf{e-h}, Ref.~\onlinecite{tosatti2001}, AAAS.}
\label{fig:magic}
\end{figure}
Predicting the actual magic shapes observed in this geometry turns out to require not just the \textit{ab initio} effort to calculate the total energy $E(N)$ (where $N$ is the number of atoms in the wire) as a function of shape, but also the thermodynamic consideration that a suspended nanowire is free to exchange atoms with the leads, actually bringing it into grand canonical equilibrium with them. The resulting equilibrium configurations are determined by the minima of
\begin{align}
f =\frac{E(N) - \mu N}{L},
\end{align}
where $\mu$ is the chemical potential of an atom in the bulk-like metal leads, and $L$ is the nanowire length.\cite{tosatti2001}  An atom in a nanowire always has a higher energy than in the infinite bulk, making $f$ positive definite.  With the dimensions of force, $f$ represents the thermodynamic {\it string tension} of the suspended nanowire, whose minimum $f=0$ is only attained for $N=0$.  The adiabatic shape evolution of a suspended nanowire in time is governed by the progressive and spontaneous decrease of $f$, sketched in Fig.~1e, actuated by jump-like drops in $N$ and therefore in radius, until the nanowire finally breaks when $N=0$.  As it turns out, the magic nanowire structures are helical (Fig.~1f versus Fig.~1g,h) because for the same number of atoms a helical twist reduces the radius and increases the length $L$, therefore lowering the nanowire string tension, very much as would happen in a rolled up sheet of paper.   Short monatomic chains of gold---the minimal nanowires---were also found and recorded in stunning real-time transmission electron microscope movies.\cite{ohnishi1998}  Break junction data clearly identified them in gold and also platinum.\cite{yanson1998,smit2001}  The simple string tension model works qualitatively in this case too.\cite{tosatti2005} 

It should be noted that the microscopic mechanisms leading to the energy minimizing structure of magic nanowires may be of different types. In noble, transition and other metals where cohesion involves a multiplicity of electronic contributions, minimal energy configurations are generally believed to correspond to optimal packing of the atoms.\cite{guelseren1998}  In alkali metals, where nanowires were observed in break-junction work,\cite{yanson2001} the energy minima may also be determined by electronic shell closing.\cite{urban2004}
\medskip

% 3
\noindent \textbf{Ballistic conductance} 

Nanocontacts operate in the ballistic, as opposed to diffusive, transport regime when the elastic mean free path is much greater than the characteristic dimension $L$ of the contact.  Quantum effects come into play when $L$ is small enough to be comparable to the Fermi wavelength $\lambda_F$.  Nanocontacts meet this requirement, as $\lambda_F$ is on the order of 5\AA~in metals.  A contact region of width $W$ is expected to support only a finite number of conducting channels $N \approx 2W/\lambda_F$ due to wave quantization in the transverse direction, analogous to waveguides.

Landauer-B\"uttiker formalism is a linear response theory that describes quantum ballistic transport in terms of the transmission and reflection coefficients of conduction electrons scattering off the contact region.\cite{landauer1957,buettiker1988,imry1999} 
A typical scattering geometry is represented in Fig.~2.  
\begin{figure}[tb!]
\includegraphics[width=0.8\columnwidth]{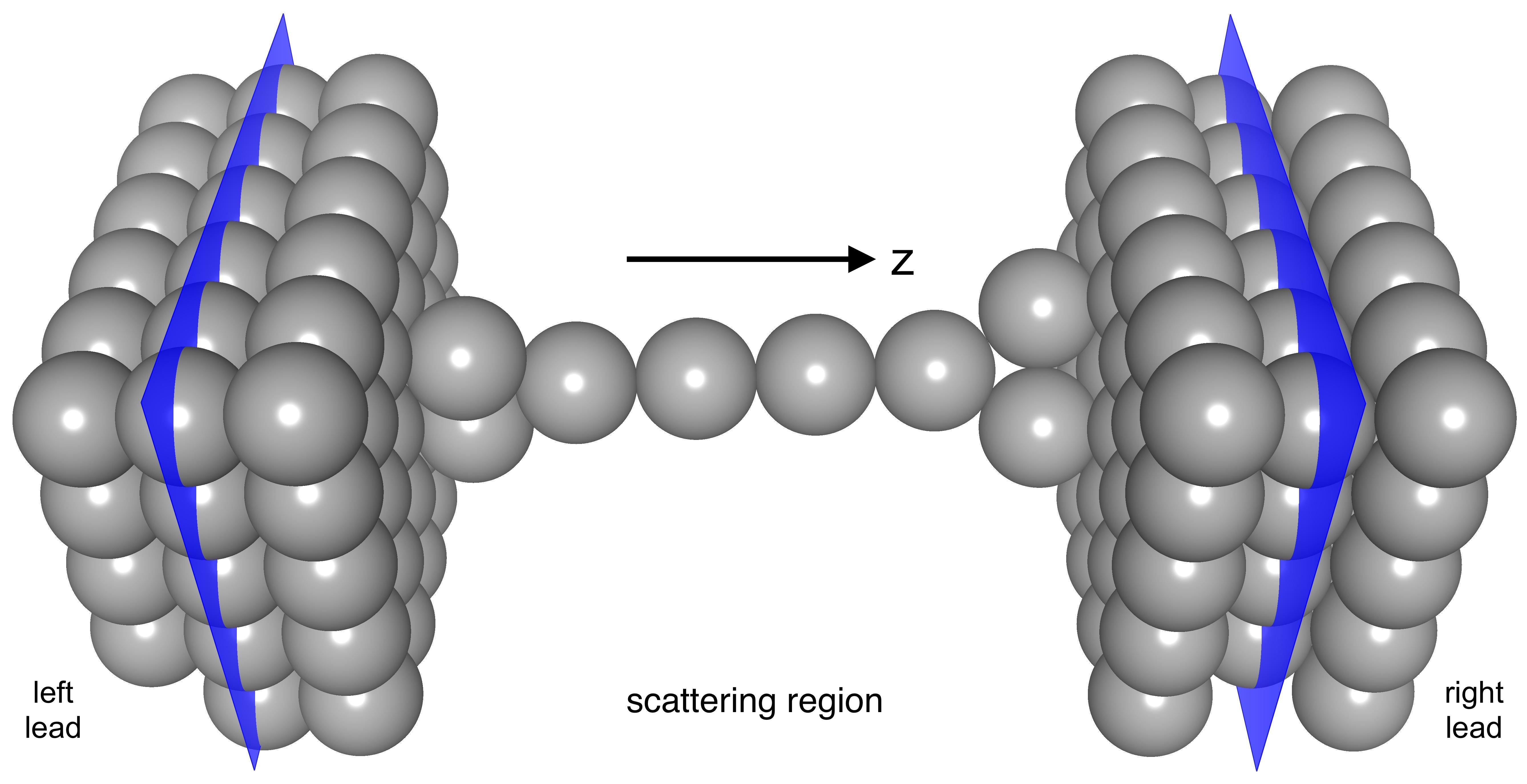}
\caption{\textbf{Model break-junction nanocontact.} Platinum nanocontact showing the scattering region and the matching planes (blue) where the potential is matched to that of the (111) bulk leads.  To achieve a smooth matching, the scattering region is chosen so as to include the entire nanocontact as well as portions of the left and right leads. Atomic coordinates are those of Ref.~\onlinecite{kumar2013}; the arrow shows the direction of the $z$ axis. Figure produced with the VESTA visualization program (http://jp-minerals.org/vesta).}
\label{fig:scattering}
\end{figure}
The number of conducting channels at a given energy $\epsilon$ in the left and right leads, $N_L$ and $N_R$, is determined by the number of bands crossing that energy.  The amplitudes of incoming and outgoing channels are related by the scattering matrix 
\begin{align}
S = \left( \begin{array}{cc} r & t' \\ t & r' \end{array} \right) {,}
\end{align}
where $t$ is the $N_R\times N_L$ matrix containing the transmission amplitudes connecting each of the $N_L$ incoming channels of the left lead with each of the $N_R$ outgoing channels of the right lead and $r$ is the $N_L\times N_L$ matrix of reflection amplitudes.  Similarly, $t'$ and $r'$ describe transmission and reflection of states incident from the right lead.  When the chemical potentials in the left and right leads are fixed to $\mu_L$ and $\mu_R$, the current through the nanocontact is given by the Landauer-B\"uttiker formula
\begin{align}
I = \frac{e}{h} \int_{-\infty}^{\infty} T(\epsilon) [f(\epsilon-\mu_{L}) - f(\epsilon-\mu_{R})] d\epsilon {.}
\label{eq:LB1}
\end{align}
Here, the energy-dependent transmission function $T(\epsilon)=\mathrm{Tr}(t^{\dag} t)$ is the trace of the product of $t$ and its hermitian conjugate $t^{\dag}$, and $f(\epsilon)$ is the Fermi-Dirac distribution function.  At zero temperature, the Landauer-B\"uttiker formula implies the differential conductance at zero bias voltage ($V$): 
\begin{align}
G = \frac{dI}{dV} = \frac{e^2}{h} T(\epsilon_F) = \frac{e^2}{h} \sum_{n\sigma} T_{n\sigma} {,}
\label{eq:LB2}
\end{align}
where the summation is over the $N_L$ eigenchannel transmission probabilities $T_{n\sigma}$ (eigenvalues of $t^{\dag} t$) and $\sigma$ is the spin index.  This remarkably simple formula depends only on the universal conductance quantum $G_0=e^2/h$ and the transmission function at the Fermi energy $T(\epsilon_F)$.  In non-magnetic nanocontacts, one often redefines the conductance quantum as $2e^2/h$ by summing over spin independent $T_n$ in Eq.~(\ref{eq:LB2}), but we will not do so here.  Quantum ballistic conductance depends on details of the quantum mechanical scattering off the contact region and is therefore a consequence of the wave nature of electrons.  Equations~(\ref{eq:LB1}) and (\ref{eq:LB2}) in fact assume that the wavefunction maintains phase coherence throughout the nanocontact region, and Eq.~(\ref{eq:LB1}) is adequate only for sufficiently low bias voltage, as the Landauer-B\"uttiker theory is based on linear response.  

Conductance quantization,\cite{agrait1993,pascual1993,krans1993,olesen1994,brandbyge1995,gai1996,ohnishi1998,yanson1998} as plotted for an STM nanocontact in Fig.~3a, is a remarkable property of quantum ballistic transport.  
%Figure 3
\begin{figure}[tb!]
\includegraphics[width=0.48\columnwidth]{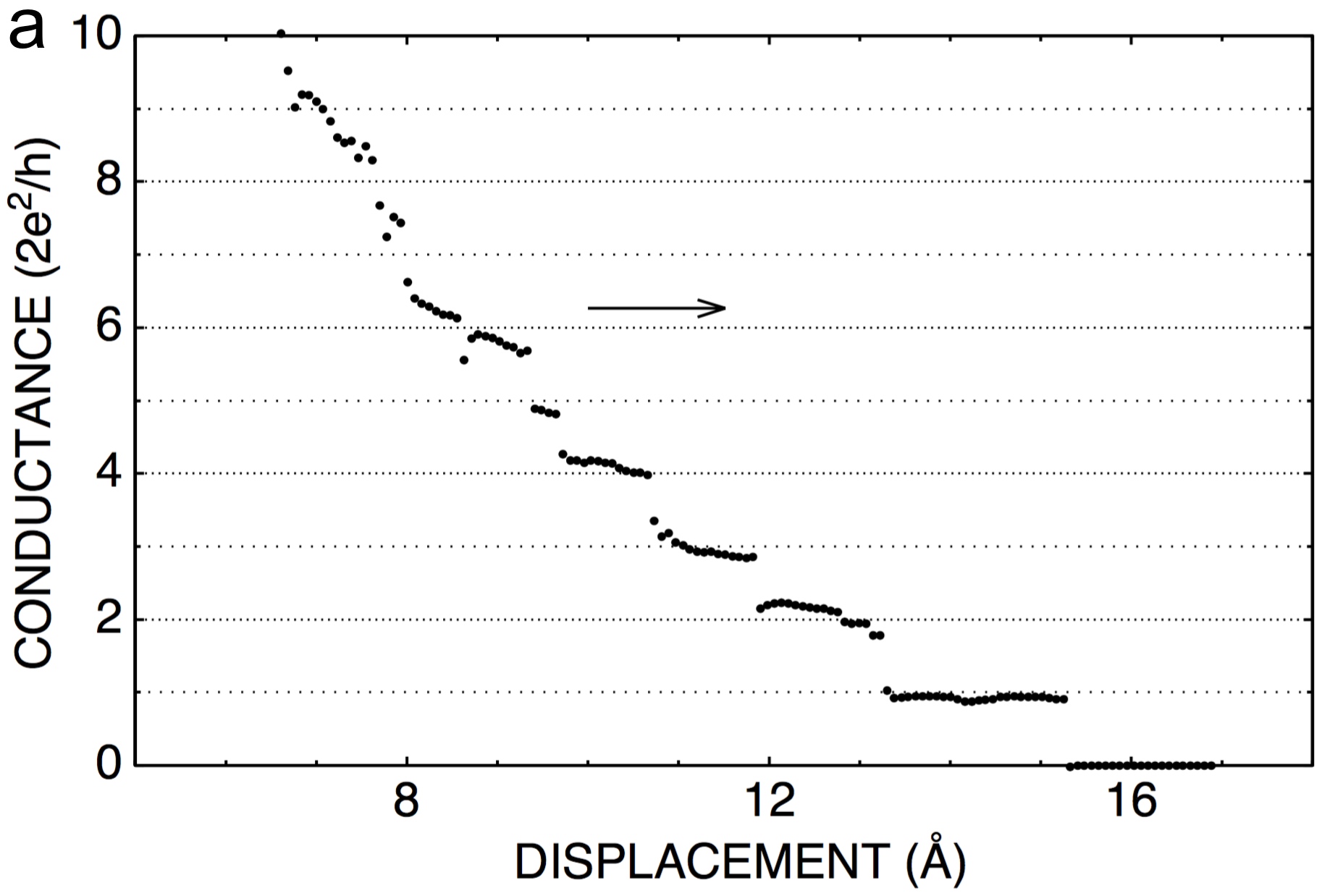}\includegraphics[width=0.48\columnwidth]{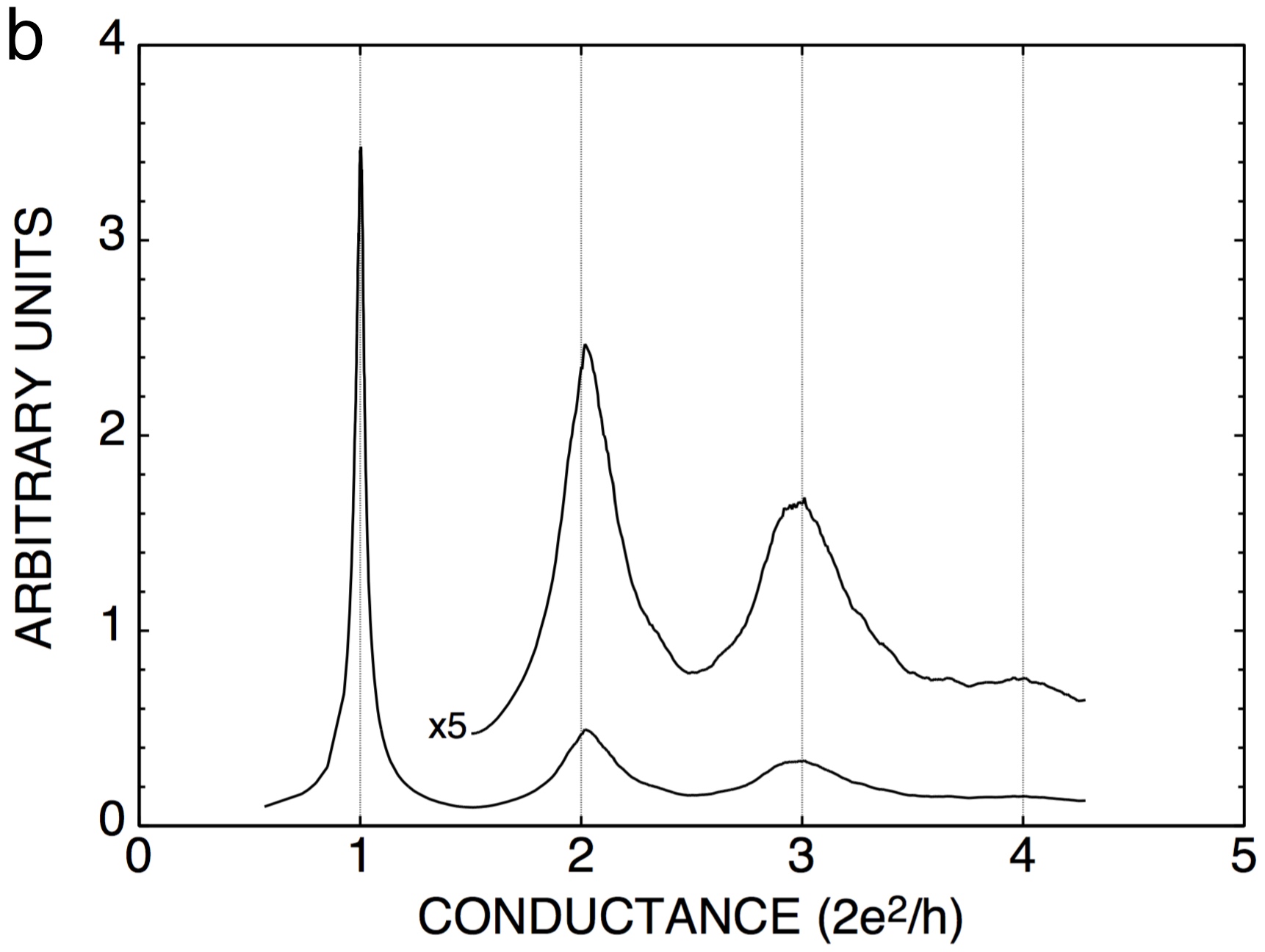}
\caption{\textbf{Conductance quantization}. \textbf{a}, Conductance of an STM nanocontact on Au(111) as a function of tip-surface displacement. When an STM tip is contacted with a Au surface and then slowly retracted (in the direction indicated by the black arrow), the conductance decreases in a stepwise fashion through a series of plateaus at approximately integer multiples of $2G_0$. \textbf{b}, Conductance histograms for Au STM nanocontacts at room temperature. The sharp peak at $2G_0$ corresponds to the monatomic chain contact of the last plateau in \textbf{a}. Figure reproduced with permission from Ref.~\onlinecite{brandbyge1995}, APS.}
\label{fig:steps:histogram} 
\end{figure}
Similar behaviour is observed in break-junction experiments.  The last plateau is particularly flat and well defined and occurs close to $2G_0$.  In some metals, the last plateau persists despite an elongation much greater than the interatomic spacing.  This indicates that the single atom contact progressively lengthens into a chain of up to $\sim$10 atoms as the tip is retracted.\cite{yanson1998,ohnishi1998,kizuka2008}  Chain formation is favoured in 5$d$ metals, such as Ir, Pt, and Au, over 4$d$ metals due to multi-atom forces,\cite{ercolessi1988,tosatti2005} ultimately rooted in relativistic effects.\cite{takeuchi1989,agrait2003}  Real-time images of the structure and mechanics of chain formation have been acquired in transmission electron microscopes.\cite{ohnishi1998,rodrigues2001,kizuka2008,lacroix2014} 
Since the width of the chain is a single atom, only one broad $s$-like conduction band is open and contributes $T_{n\sigma}\approx 1$ per spin, while all other bands have $T_{n\sigma}\approx 0$.  For other metals, particularly transition metals with partially filled $d$ bands, the lowest plateau often occurs at a conductance greater than $2G_0$.   An upper limit is set by the chemical valence of the bridging atom.\cite{scheer1998}  

The atomic-scale variability of the nanocontacts causes a degree of randomness in the positions of the conductance plateaus, but after several repetitions of the experiment a preference for certain values can be discerned.  To analyse this trend systematically, large data sets of individual conductance versus displacement curves are collected and plotted as histograms.  Figure 3b shows an example of a histogram for Au nanocontacts at room temperature.\cite{brandbyge1995}  Similar histograms are obtained for other noble metals (Cu, Ag) and the alkali metals (Li, Na, K).  \textit{Ab initio} calculations for short Au chains suspended between bulk leads\cite{brandbyge2002,palacios2002,mehrez2002} reproduce well the experimental conductance 2$G_0$ of the last plateau.

Although many of the lowest conductance plateaus occur close to integer multiples of $2G_0$, it is not straightforward to interpret them as evidence of conductance quantization of the type found in two-dimensional electron gas point contacts,\cite{vanwees1988,wharam1988} where the conductance jumps sharply but continuously by $2G_0$ as the width of the contact is varied, because the cross-section of a nanocontact cannot be varied continuously, due to the atomic granularity.\cite{agrait2003}  From simultaneous measurements of the conductance and  force on the STM tip,\cite{rubio1996,rubiobollinger2001} as well as analogous molecular dynamics simulations,\cite{todorov1996,sorensen1998} the sharp jumps in the conductance have been correlated with abrupt atomic-scale rearrangements of the nanocontact geometry.  The height of the conductance steps is not always quantized to an integer multiple of $2G_0$, and their position varies from trace to trace and from metal to metal.  Thus, it is not always straightforward to attribute them to the opening or closing of individual channels with $T_{n\sigma}$ close to 1.

%    ---    ab initio calculations of ballistic conductance    ---

The key input of the Landauer-B\"uttiker formula is the scattering matrix $S$, but the formalism itself does not provide a prescription for modelling the nanocontact geometry or calculating its $S$ matrix.  Many different approaches have been developed for that purpose (see Refs.~\onlinecite{brandbyge2002,palacios2002,calzolari2004} and references therein).  Here we review an \textit{ab initio} method for solving the quantum mechanical scattering problem for realistic nanocontacts,\cite{smogunov2004a} based on the complex band structure approach.\cite{choi1999}  An implementation of the method is available in the electronic structure code Quantum Espresso.\cite{giannozzi2009}

The self-consistent Kohn-Sham potential from a ground-state density functional theory (DFT) calculation of the scattering region and left and right leads, as illustrated in Fig.~2, provides the potential defining the quantum mechanical scattering problem.  The asymptotic form for the scattering state $|\Psi\rangle$ of the $j$th incoming Bloch wave $|\psi_j\rangle$ of the left electrode as $|z|\rightarrow \infty$ is 
\begin{align}
|\Psi\rangle \rightarrow \left\{ \begin{array}{ll} 
|\psi_j\rangle + \sum_{i\in L} r_{ij} |\psi_i\rangle, & \textrm{in the left (L) lead} \\
\sum_{i\in R} t_{ij} |\psi_i\rangle, & \textrm{in the right (R) lead} \end{array} \right. {.}
\end{align}
The wavefunction in the scattering region is matched to the wavefunctions in the leads by requiring continuity of $\Psi(\mathbf{r})$ and $\partial\Psi(\mathbf{r})/\partial z$ at the matching planes shown in Fig.~2.  To achieve a smooth matching, it is crucial that the wavefunction in the left lead is expanded not only over the propagating Bloch waves with real crystal momentum $k_z$ but also over generalized Bloch states with \textit{complex} $k_z$, which decay to the left.  Analogous considerations apply to the right lead.  Inside the scattering region, where the translational symmetry is broken in the $z$ direction, the Schr\"odinger equation is solved numerically on a dense sequence of planar slices. 

This approach delivers the full $S$ matrix and the scattering phase shifts $\delta_j$, defined from the eigenvalues of $S$, diagonalized by a unitary matrix $U$, i.e.
\begin{align}
U^{\dag} S U  = \left( \begin{array}{ccc} 
e^{i2\delta_1} & 0 & 0   \\
0 & e^{i2\delta_2} & 0  \\
0 & 0 & \ddots  \end{array} \right) {.} 
\end{align}
The scattering phase shifts, which depend on the energy of the incoming wave, can be used to calculate several relevant quantities, such as the orbital-resolved density of states of a given scattering eigenchannel.  The approach includes full spin-orbit interactions, noncollinear magnetism and, if relevant, corrections to the nanocontact electronic structure due to the Hubbard repulsion and exchange coupling of localized orbitals.  Any atomistic model can be used for the leads and scattering region, which are treated on equal footing.  Inelastic vibrational or magnetic excitations and non-equilibrium effects are not directly included.
\medskip

% 4
\noindent \textbf{Magnetic junctions} 

In nanocontacts with ferromagnetic leads, one might expect conductance quantization in units of $e^2/h$ (instead of $2e^2/h$), if spin degeneracy is lifted.  This is not what is observed\cite{untiedt2004} because in magnetic nanocontacts there are typically several partially open $d$ channels which generally add up to give non-integer $G/G_0$.  Moreover, there is no peak at $e^2/h$, as the occurrence of magnetism in the $d$ bands does not shut off the $s$-like minority spin channel,\cite{smogunov2004a} except in the presence of impurities.\cite{untiedt2004} 

Histograms of STM conductance measurements show that Ni has broad peaks at 2.6$G_0$-3.2$G_0$ and 6$G_0$ (Refs.~\onlinecite{sirvent1996,oshima1998,ono1999,bakker2002,untiedt2004}),  Co has broad peaks at 2.5$G_0$ and 5$G_0$ (Ref.~\onlinecite{untiedt2004}) and Fe has a broad peak at 3.6$G_0$-4.4$G_0$ (Refs.~\onlinecite{ludoph2000,untiedt2004}).  The conductance of the lowest peak (the last plateau before breaking) is thus much smaller than the upper bound of $\sim 8G_0$ based on the valency of the contact atom.  Although the transmittance of light-mass $s$ channels remains close to 1, that of heavy-mass $d$ channels is much smaller.  Despite this complexity, in addition to that associated with uncertainties in the structure and magnetic state of the contact,
it has nevertheless proved possible to achieve semi-quantitative agreement between \textit{ab initio} conductance calculations and experiment for a wide range of metals.  

\textit{Ab initio} calculations for a Ni nanocontact formed by two end-on (001) pyramid-shaped leads whose vertex atoms are 2.6~\AA~apart found a conductance of 3.85$G_0$; the minority spin channel contributed 2.8$G_0$ and the majority spin channel 1.05$G_0$ (Ref.~\onlinecite{jacob2005}). Calculations for a three-atom Ni chain suspended between bulk (001) electrodes using the \textit{ab initio} Landauer-B\"uttiker approach described in the previous section obtained $3.2G_0$, of which $2.2G_0$ comes from the minority spin channel and 1.0$G_0$ from the majority channel.\cite{smogunov2006}  The conductance of linear and zigzag three-atom Co chains between (001) Co leads was calculated to be $1.3G_0$-$1.8G_0$, depending on the geometry and magnetic configuration.\cite{bagrets2004}

The magnetization of the left and right ferromagnetic leads is not necessarily aligned in experiment.  In calculations of an ideal monatomic Ni nanowire, the presence of a magnetic domain wall, that is, antiparallel magnetization of the two leads, has been shown to completely block the $d$ channels, lowering the conductance from the ideal value of 7$G_0$ to 2$G_0$.\cite{smogunov2002,smogunov2004a,smogunov2004b} 
This results in very high values of magnetoresistance, MR$=(G_P-G_{AP})/G_{AP}\times 100\%=250\%$, where $G_P$ and $G_{AP}$ are the conductances for parallel and anti-parallel magnetic alignments.  For more realistic two-atom Ni nanocontact geometries, more moderate values of MR~$\approx 30\%$ were theoretically reported,\cite{jacob2005} as a consequence of the much smaller conductance of the $d$ channels due to their strong contact back-scattering.  Similarly, the MR of few-atom chains of Co, Cu, Al and Si suspended between Co leads was found to be below 50\% in all cases investigated.\cite{bagrets2004} Another important property is the anisotropic magnetoresistance, which reflects the dependence of the conductance on the relative angle between the electric current and the magnetization direction.  This phenomenon was reported in the ballistic regime for Ni, Co, and Fe atomic-scale nanojunctions,\cite{autes2008,sokolov2007} following a theoretical prediction.\cite{velev2005} 

The degree of spin-polarisation of the current, defined as 
$(G_\uparrow-G_\downarrow)/(G_\uparrow+G_\downarrow)\times 100\%$, where $G_\sigma=(e^2/h) \sum_n T_{n\sigma}$ is the total spin-$\sigma$ conductance, is restricted in magnetic transition metal nanocontacts by the fact that both spin-up and spin-down $s$-like channels conduct almost perfectly.  Therefore, the total current is only partially spin-polarized, e.g., values of about $33\%$ were reported in Ni nanocontacts.\cite{smogunov2006}  A symmetry-based mechanism to fully block the spin-up conductance by joining two ferromagnetic electrodes with a special class of nonmagnetic $\pi$-conjugated molecule has recently been proposed.\cite{smogunov2015}  Molecules with only $\pi$-symmetry orbitals around the Fermi level will reflect both spin-up and spin-down $s$-symmetry channels coming from the ferromagnetic electrode.  Only spin-down $d$ channels compatible with the molecular $\pi$-orbitals will be allowed to pass through the junction, providing perfect spin-polarized conductance.   A similar blocking mechanism was predicted to occur due to the $\pi$-symmetry $p$ orbitals of oxygen impurities in Ni nanocontacts, leading to a MR of $\sim$700\%.\cite{jacob2006}

Magnetism was predicted to emerge spontaneously in reduced dimensions in the nominally non-magnetic metals platinum\cite{delin2003} and palladium.\cite{gava2010}  In DFT calculations, a Pt nanowire chain magnetizes parallel to the direction of the wire (easy axis magnetization) as a result of the narrowing of the $d$ bands and the cooperative effect of spin-orbit-coupling induced orbital magnetic moments.\cite{delin2003,delin2004b,smogunov2008a,smogunov2008b}  The ballistic conductance of a one-atom Pt contact suspended between (001) bulk leads was calculated to be 2.1$G_0$,\cite{fernandezrossier2005} and for three-, four- and five-atom Pt chains it varies between 2.0$G_0$ and 2.3$G_0$,\cite{smogunov2008a} in good agreement with break-junction experiments.\cite{krans1993,smit2001,smit2002,nielsen2003}  The calculations have also revealed rather moderate values of ballistic anisotropic magnetoresistance, up to $14\%$ for the longest five-atom Pt nanocontacts.\cite{smogunov2008a}  Magnetism in Pt break-junction nanocontacts with approximately one- to four-atom chains has recently been confirmed by a gamut of experimental measurements.\cite{strigl2015}
\medskip

% 5
\noindent \textbf{Shot noise and contact magnetism} 

Shot noise is a measure of non-equilibrium current fluctuations.  While ballistic conductance is connected with the wavelike behaviour of electrons, shot noise depends on their particle nature through the discreteness of the electron charge.  Assuming a Poissonian distribution of electron arrival times (tunnelling events), Schottky showed that the current noise power spectrum $S(\omega)$ is approximately frequency independent and derived the classical relationship $S=2e\langle I\rangle$ where $\langle I\rangle$ is the average current.  The current noise power spectrum is defined as 
\begin{align}
S_{ab}(\omega) = \frac{1}{2} \int e^{i\omega t} \langle \Delta \hat{I}_a(t) \Delta \hat{I}_b(0) + \Delta \hat{I}_b(0) \Delta \hat{I}_a(t) \rangle dt {,}
\end{align}
where $\Delta \hat{I}_{\alpha} = \hat{I}_{\alpha} - \langle \hat{I}_{\alpha} \rangle$ and $\hat{I}_{\alpha}$ is the current in lead $\alpha$.  In the low-temperature limit $k_B T \ll eV$ (where $k_B$ is the Boltzmann constant and $T$ is temperature) and with the help of approximations relevant for ballistic nanocontacts (see Ref.~\onlinecite{blanter2000} for a review), the noise power at a given lead $a$ simplifies to $S_{aa} = 2eF \langle I\rangle$, where the Fano factor
\begin{align}
F = \frac{\sum_{n\sigma} T_{n\sigma} (1-T_{n\sigma})}{\sum_{n\sigma} T_{n\sigma}} \label{eq:fano}
\end{align}
describes the quantum mechanical suppression of Schottky's classical result. 

Since shot noise depends nonlinearly on the transmission probabilities $T_{n\sigma}$, it represents an independent experimental probe providing information about the $T_{n\sigma}$ beyond that obtained from ballistic conductance measurements.  The quantized conductance steps observed during the stretching of a Au nanocontact have been attributed to the spontaneous closing of channels with each successive atomic rearrangement of the local contact geometry.  All $T_{n\sigma}$ are presumed to be either fully open or fully closed (1 or 0), except a single spin-degenerate channel carrying a fraction of a conductance quantum, i.e.,~$0\leq T_{n\sigma}\leq 1$.  If this is the case, then according to Eq.~(\ref{eq:fano}) the shot noise should be suppressed when $G=2n G_0$; $n=0,1,2,\ldots$  This suppression is indeed observed,\cite{vandenbrom1999} corroborating the standard interpretation.  In other metals such as Al, the measured shot noise is not consistent with this scenario and can only be explained if multiple fractionally-open channels contribute to the conductance, \cite{vandenbrom1999} which agrees with the results of independent measurements on superconducting Al nanocontacts.\cite{scheer1997}  Recent investigations of STM Au break junctions have found that also the ensemble variance of the shot noise $\langle S^2 \rangle - \langle S\rangle^2$ is suppressed at $G\approx 2nG_0$ (Ref.~\onlinecite{chen2014}).

For a given value of the conductance $G$, the shot noise has definite maximum and minimum values corresponding to different distributions of the $T_{n\sigma}$. 
The upper bound is achieved by taking all $T_{n\sigma}=0.5$.  The lower bound occurs when all channels except one are either fully open or fully closed, as occurs in Au nanocontacts.  Since the lower bound for magnetic states is below the lower bound of non-magnetic states, shot noise may be used as an experimental probe of nanocontact magnetism.\cite{kumar2013,burtzlaff2015}  Simultaneous measurements of $G$ and $F$ for elongated Pt nanocontacts found a cluster of points all lying above the non-magnetic lower bound with a significant accumulation of points right at the boundary, as shown in Fig.~4.  
% Figure 4
\begin{figure}[tb!]
\includegraphics[width=0.8\columnwidth]{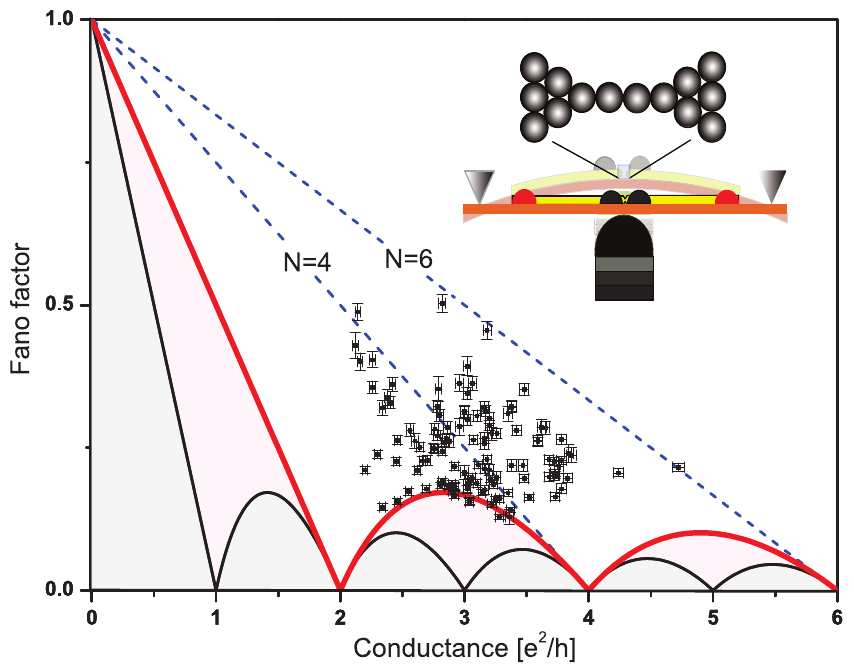}
\caption{\textbf{Shot noise measurements.} Scatter plot of the conductance $G$ and Fano factor $F$ for 119 different Pt atomic chain nanocontacts.  Black (red) curve shows the lower bound of the magnetic (non-magnetic) region.  A violation of the non-magnetic lower bound would constitute proof of magnetism.  The dashed blue lines show the maximum possible noise for a nanocontact with $N=4$ and $N=6$ channels.  Inset: Illustration of a mechanical break junction actuated by a pushing rod.  Figure reproduced with permission from Ref.~\onlinecite{kumar2013}, APS.}
\label{fig:noise}
\end{figure}
While this at first suggests a non-magnetic state, \textit{ab initio} ballistic transport calculations for the nanocontact geometry in Fig.~2 
determined that the results are nevertheless consistent with a magnetic state, as the $s$ channels involved in conductance, and hence in shot noise, are weakly spin-polarized ``spectators'' of the magnetic state of the nanocontact, whereas the ``actor" $d$ channels driving magnetism are very poorly conducting.\cite{kumar2013}  Simultaneous $G$ and $F$ measurements have been used to estimate the spin polarisation of the conductance through Co and Fe adatoms on the Au(111) surface.\cite{burtzlaff2015}
\medskip

% 6
\noindent \textbf{Kondo conductance anomalies} 

The nanocontact magnetism reviewed thus far has been static magnetism, a situation where, as in bulk magnets, the symmetry breaking mean-field picture of local magnetic moments aligned in a specific direction is valid and one can talk about separate spin-up and spin-down conductance channels.  This picture breaks down if the magnetic moment of the contact region is small enough that it can flip through spin-flip scattering with conduction electrons on the experimentally relevant timescale, determined by the inverse width of the sharpest zero bias features in the I-V spectrum.  In such situations the magnetism is dynamical.

In the simplest case of a spin-$\frac{1}{2}$ single atom nanocontact, spin-flip scattering ``screens'' the local magnetic moment, creating a many-body singlet state and restoring the $SU(2)$ spin symmetry which was broken in a static mean-field calculation.  This is the Kondo effect in a nutshell.\cite{hewson1993}  The screening is gradually suppressed as the temperature is raised above the characteristic Kondo temperature $T_K$, which depends on the strength of the coupling with conduction electrons.  Relatively difficult to detect experimentally in bulk materials, the Kondo effect is utterly relevant to transport in magnetic nanocontacts because the low-energy spin-flip scattering processes generate a resonance in the density of states at the Fermi energy, fundamentally altering low-bias transport.  For typical experimentally measurable $T_K$ in the range of 1-100~K, the Kondo screening occurs on a short timescale $\hbar/k_B T_K\approx 10^{-13}-10^{-11}$~s, meaning that transport experiments see the many-body singlet state and not the bare impurity spin.  The Kondo effect has been observed in scanning tunnelling spectroscopy of Co adatoms on gold,\cite{madhavan1998} silver,\cite{wahl2004} and copper\cite{wahl2004,manoharan2000,neel2007,vitali2008} surfaces, Ce adatoms on silver\cite{li1998} and Fe and Co impurities beneath the copper surface,\cite{prueser2011} and in Fe, Co, and Ni break junctions\cite{calvo2009} as a sharp zero-bias conductance anomaly with Fano lineshape.\cite{ujsaghy2000} 

The Kondo effect is also relevant for nanocontacts containing magnetic molecules,\cite{scott2010} for example, molecules containing transition metals or radicals with unpaired electrons.  Essentially, the Kondo anomaly at the Fermi energy provides a transparent path at zero bias through molecules that would otherwise be insulating and totally reflecting.  Zero-bias Kondo conductance anomalies have been reported in break-junction experiments with transition metal-bearing molecules,\cite{park2002,liang2002,parks2010,rakhmilevitch2014} $C_{60}$ molecules,\cite{yu2004,parks2007,roch2009} a long $\pi$-conjugated chain\cite{osorio2007} and an organic radical,\cite{frisenda2015} and scanning tunnelling spectroscopy of adsorbed transition metal-bearing molecules,\cite{wahl2005,zhao2005,iancu2006,komeda2011,mugarza2011,minamitani2012,karan2015} a $\pi$-conjugated molecule,\cite{temirov2008} and organic radicals.\cite{muellegger2013,zhang2013,requist2014} 
% Figure 5
\begin{figure}[t!]
\includegraphics[width=0.68\columnwidth]{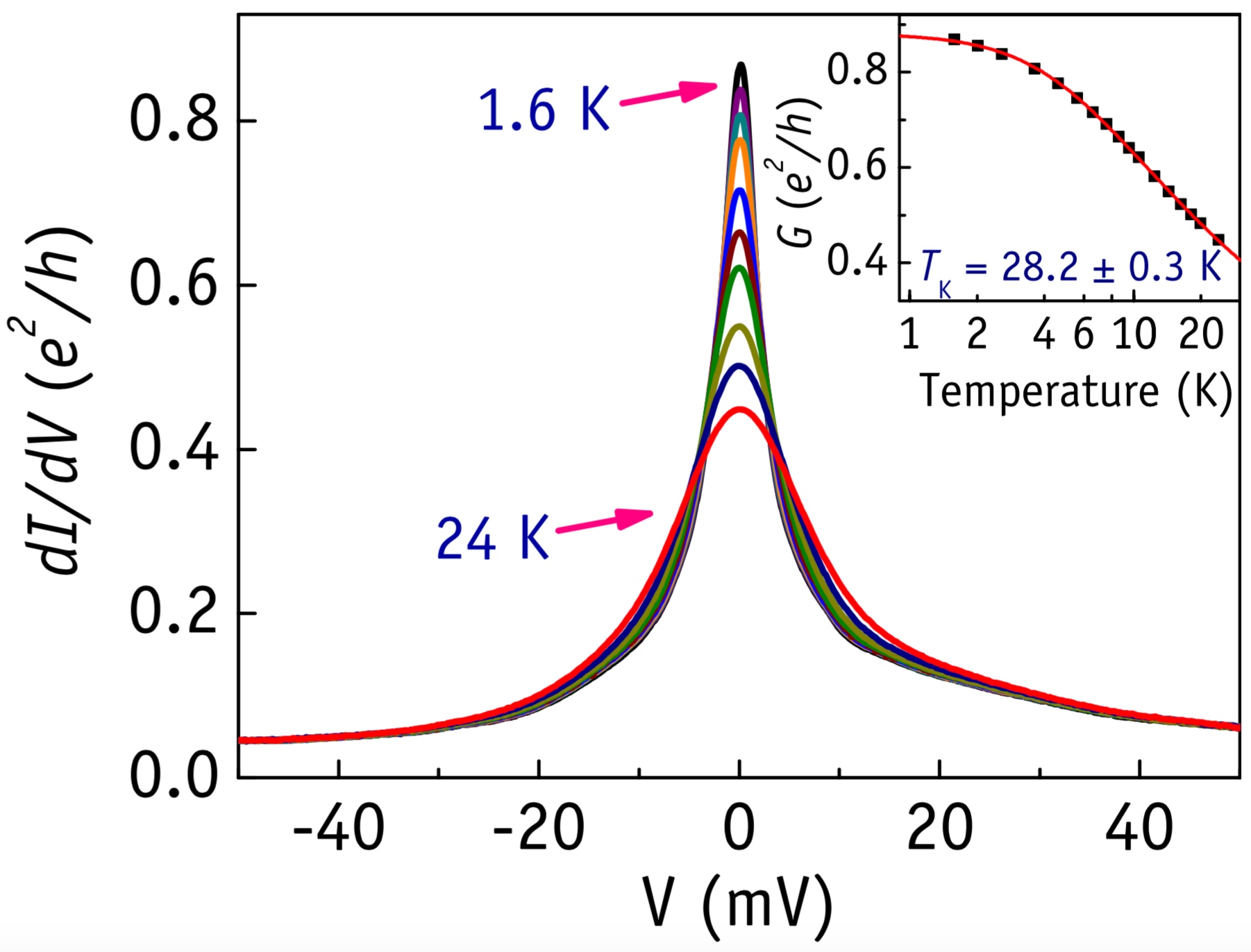}
\caption{\textbf{Kondo conductance anomaly.} Kondo zero-bias anomaly measured for a $C_{60}$ molecule in a break junction.  The suppression of the resonance with increasing temperature is a characteristic signature of the Kondo effect, which, together with its magnetic-field dependence, is used to verify the Kondo origin of zero-bias anomalies.  The inset shows a fit of the conductance to an NRG interpolation function.  Figure reproduced with permission from Ref.~\onlinecite{parks2007}, APS.}
\label{fig:kondo}
\end{figure}
Figure~5 shows an example of a zero-bias anomaly and its temperature dependence for a spin-$\frac{1}{2}$ Kondo system.

Break junction-like nanocontacts containing a suspended molecule can also be created in an STM as illustrated in Fig.~6.
% Figure 6  
\begin{figure}[tb!]
\includegraphics[width=0.7\columnwidth]{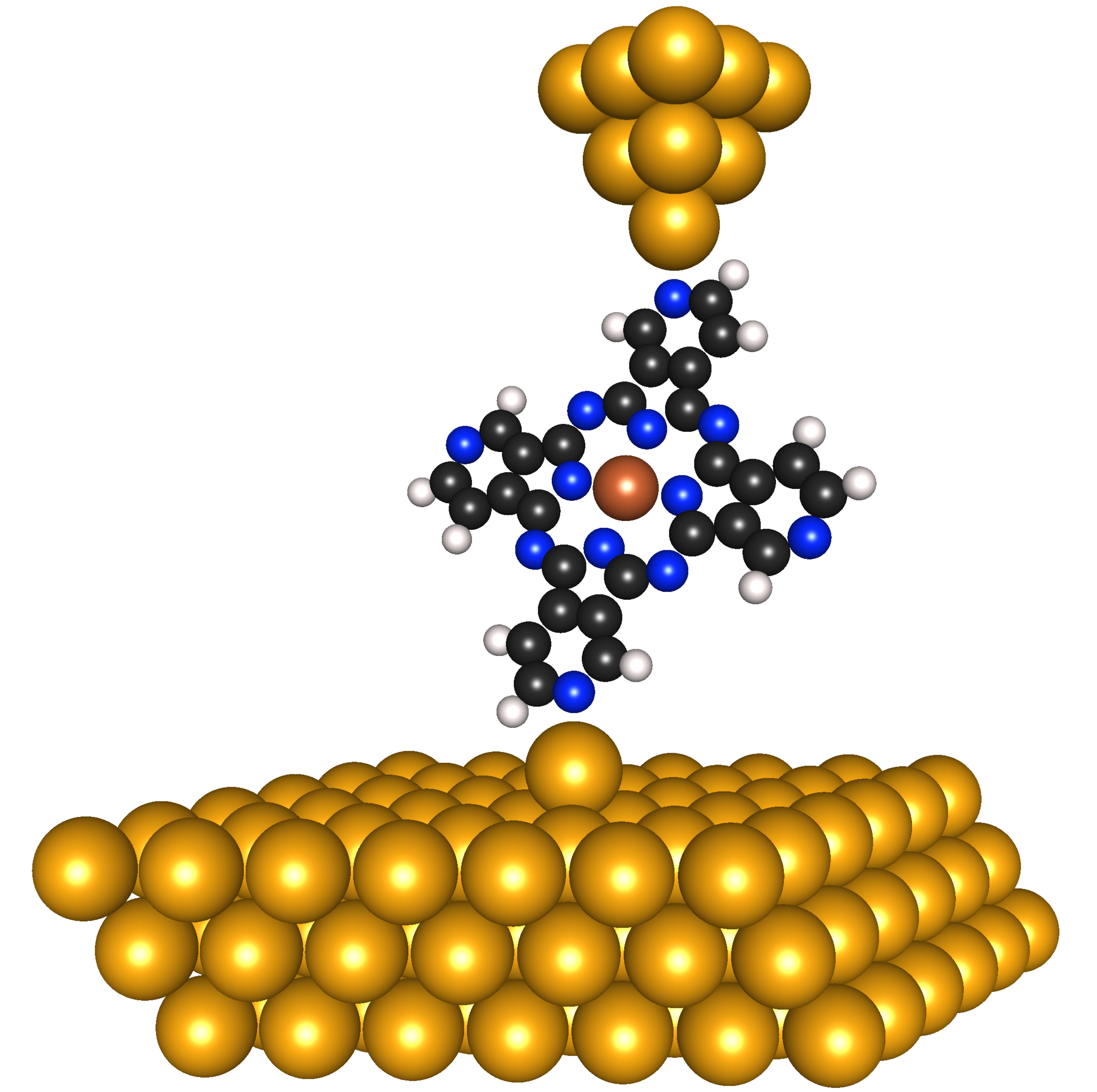}
\caption{\textbf{Scanning tunnelling microscope molecular break junction.}  A hypothetical break-junction-like molecular nanocontact consisting of a Cu tetraaza-phthalocyanine molecule suspended between an STM tip and Au(111) surface is illustrated.  Pyridine side groups provide bonding to Au.  Such molecular nanocontacts have been prepared by pressing an STM tip into the surface near an adsorbed molecule.\cite{xu2003,quek2009,aradhya2013} When the tip is retracted, there is a certain probability that one part of a molecule will bond to the tip and lift off the surface.  Colors: (Au, Cu, C, N, H) = (gold, copper, black, blue, white).  Figure produced with the VESTA visualization program (http://jp-minerals.org/vesta).}
\label{fig:pickup}
\end{figure}
The molecule usually contains chemical linker groups that bond to metals, such as thiols, amines or pyridines.  As in break junctions, the nanocontact can be mechanically controlled by varying the distance between the tip and surface.\cite{xu2003,quek2009,aradhya2013}  Molecular contacts between an STM tip and surface have also been found to form spontaneously\cite{haiss2003} or after an applied voltage pulse.\cite{sotthewes2014}

The type of Kondo effect (fully screened, underscreened, singlet-triplet, two-stage, etc.) that occurs in a given nanocontact is totally determined by the nature of the conducting channels, which are in turn dependent on the local atomic positions and magnetic polarisation.  In calculations, those details are generally decided on the basis of Anderson impurity model calculations with empirically chosen parameters -- an approach which becomes increasingly difficult in multiorbital, multivalence situations and one which does not unambiguously determine the atomic or molecular orbitals involved.  \textit{Ab initio} calculations single out the relevant localized states and interactions but cannot be used to directly calculate zero-bias conductance anomalies since the static mean-field picture of spin-polarized DFT is not valid for dynamical magnetism and the common semi-local exchange-correlation functionals do not include Kondo correlations.

The \textit{ab initio} ballistic scattering approach described above has been extended\cite{lucignano2009} to treat Kondo conductance anomalies by linking DFT with the numerical renormalization group method (NRG),\cite{hewson1993,bulla2008,zitko2009} a many-body technique for calculating the Green's function and thermodynamic properties of localized magnetic states in contact with conduction bands.  The DFT+NRG method was originally used to calculate the Kondo conductance of an idealized Au nanowire with Ni impurity.\cite{lucignano2009}  Calculations have also reproduced experimental trends in the Kondo temperature of a Co adatom on different Cu surfaces,\cite{baruselli2015} although discrepancies remain for the Fano lineshapes, which are sensitive to surface states and the atomic structure of the tip-adatom-surface contact.  Being a parameter-free method, the DFT+NRG method has been used to predict Kondo conductance anomalies for impurities on carbon nanotubes.\cite{baruselli2012b}  The Kondo anomaly in the simplest molecular radical, nitric oxide, on the Au(111) surface was predicted in this way and subsequently confirmed by scanning tunnelling spectroscopy.\cite{requist2014}  NRG calculations based on effective models have also been made for larger molecules.\cite{diasdasilva2009,cornaglia2010}  Non-equilibrium effects, such as those occurring at finite bias, are potentially relevant in some situations but are not covered in this Review.

Although NRG is a numerically exact ``gold standard,'' it can only correlate two or three orbitals at a time due to computational limits.  Other approaches to building many body correlations on top of mean-field DFT calculations\cite{jacob2009,costi2009,surer2012} are able to approximately correlate the entire $d$ shell of a transition metal impurity.
Calculations have been made for transition metals in metallic hosts\cite{surer2012} and large surface-adsorbed inorganic molecules, such as manganese phthalocyanine\cite{jacob2013} and manganese porphyrin.\cite{karan2015}
\medskip

% 7
\noindent \textbf{Exotic Kondo effects} 

In the previous section we mainly discussed fully screened Kondo models, which show Fermi-liquid properties {\it \'a la} Nozi\'eres;\cite{nozieres1974} in these systems a spin $N/2$ is screened by $M$ conduction channels, with $N=M$.  Models displaying non-Fermi-liquid behaviour include under- and over-screened Kondo models,\cite{nozieres1980,hewson1993} where, respectively, $N>M$ and $N<M$; when $N=1$ and $M=2$, the problem is usually known as the two-channel Kondo effect.  Even though these models are difficult to achieve in bulk magnetic alloys, where the metal hosts possess enough scattering channels to perfectly screen the magnetic impurity ($N=M$), they can be realized in nanocontacts.  For example, the underscreened Kondo effect has been observed for spin-1 Co complexes\cite{parks2010} and C$_{60}$ molecules\cite{roch2009} in break junctions.

% FKM THEORY
One case which so far remains elusive is the ferromagnetic Kondo model,\cite{koller2005} where $N=M$ but the impurity and the conduction electrons are coupled ferromagnetically by an effective exchange interaction $J_\text{eff}$ ($J_\text{eff}<0$) rather than antiferromagnetically ($J_\text{eff}>0$) in the Kondo exchange Hamiltonian $J_\text{eff}\: \mathbf{s}\cdot\mathbf{S}$ with spin operators $\mathbf{s}$ and $\mathbf{S}$ for the conduction channel and impurity. The ferromagnetic Kondo effect will occur, for instance, when the impurity contains one level which is spin-polarized but totally uncoupled to the leads, and another level that couples to the leads but is only weakly spin polarized by intra-site Hund's coupling.  Although examples of this are at present only theoretical,\cite{koller2005,mehta2005,lucignano2009} the ferromagnetic Kondo model is important as the simplest example of non-Fermi liquid behaviour.  At low temperature the impurity spin behaves essentially as a free local moment, apart from logarithmic singularities,\cite{koller2005,mehta2005} since a ferromagnetic coupling renormalizes to zero.\cite{anderson1970}  Its spectral function displays  a logarithmic ``dimple" at $\omega =0$ and has peculiar temperature and magnetic-field dependence.\cite{baruselli2013}  A finite temperature $T$ cuts off the logarithmic dimple at low frequency, while a magnetic field, no matter how small, destroys the logarithmic dimple replacing it with a symmetric pair of inelastic spin-flip Zeeman excitations.

% FKM PROPOSALS AND EXPERIMENTS
Even though an experimental demonstration of the ferromagnetic Kondo effect is still lacking, a few proposals have been put forward, including $3d$ transition metal impurities in the bulk of $4d$ transition metals,\cite{gentile2009} substitutional Ni atoms in Au nanocontacts under strain,\cite{lucignano2009} and systems of multiple laterally coupled quantum dots. \cite{kuzmenko2006,baruselli2013,mitchell2013,andrade2015}  In this last case, it is possible to induce a Berezinskii-Kosterlitz-Thouless transition through application of a gate potential or an additional interdot coupling.\cite{baruselli2013,mitchell2013} Ferromagnetic Kondo systems could in principle be realized with surface-adsorbed molecular radicals; however, up to now, only radicals with small, but still positive $J_{\text{eff}}$, have been observed.\cite{zhang2013}
\medskip

% 8
\noindent \textbf{Technological potential} 

Representing the smallest foreseeable interface between nanoscale electronic devices and metallic electrodes, the nanocontacts reviewed here are likely to be an integral part of future electronic technologies.  In molecular devices with metallic leads, it will be crucial to understand the electronic structure of the electrode-molecule bond.  We now survey some specific technologically relevant applications involving nanocontacts.

% --molecular electronics--
After the seminal proposal of Aviram and Ratner\cite{aviram1974} envisioning a donor-acceptor molecule functioning as a diode, many other molecular devices have been proposed; for reviews of experimental progress, see e.g.~Refs.~\onlinecite{joachim2000,tao2006,heath2009}.  The conductance through molecular diodes, transistors and other devices can be calculated with the \textit{ab initio} transport theories reviewed here.

% --molecular diodes and transistors--
Current rectification has been realized in several different molecular nanocontacts,\cite{metzger2003,elbing2005,diezperez2009,batra2013} where molecular orbital-electrode hybridisation is thought to crucially influence diode performance.  Realising molecular transistors poses further challenges because they require a method of gating the source to drain current.  Single-molecule break-junction devices have been gated from below via the substrate,\cite{park2000,park2002,liang2002,yu2004,roch2008,song2009} which proves the possibility of electrically gating molecular orbitals, even if not yet providing a scalable solution for molecular electronics.  STM molecular nanocontacts are not conducive to electrical gating by a third terminal; however, the electrostatic potential of nearby charged adatoms can provide a gate potential.\cite{martinezblanco2015}  Other ways to achieve gating have been proposed, including mechanical, chemical, optical, magnetic and electrochemical gates (see Ref.~\onlinecite{tao2006} and references therein).  Mechanical compression of an STM bridge molecule has also been demonstrated to modulate conductance.\cite{quek2009,sotthewes2014}  

% --spintronics--
Computing devices that would make use of the electron spin attract interest for their advantages in terms of efficiency, storage density and the potential unification of memory and logic operations in the same device.  We highlight here a few applications where ballistic conductance through magnetic and molecular nanocontacts may be of relevance to spintronics applications.
  
% --magnetoresistance--
Memory devices such as magnetic random access memory and hard-disk read heads rely on the magnetoresistance of two ferromagnetic plates separated by a thin insulating layer.  Attempts to scale down the basic memory unit by using atomic and molecular nanocontacts with high magnetoresistance, that is,``spin valve'' behaviour, are being pursued.  Generally, only modest values of magnetoresistance are obtained in metallic nanocontacts,\cite{bagrets2004,jacob2005,smogunov2008a} although a larger value of $73\%$ has been calculated for a Au chain between Co electrodes, \cite{sivkov2014} a configuration which also realizes a relatively highly spin-polarized conductance due to the blocking of the minority spin channels by strong $s$-$d_{z^2}$ hybridisation at the Co/Au interface.  Larger values of magnetoresistance have been calculated and observed for molecular nanocontacts,\cite{pati2003,rocha2005,schmaus2011,tao2013,zu2013,smogunov2015} reaching (ideally) infinite magnetoresistance for certain non-magnetic $\pi$-conjugated molecules between ferromagnetic leads.\cite{smogunov2015}

% --spin filters--
Highly selective spin filters are sought for spintronics applications.  As reviewed in the \textsf{Magnetic Junctions} section, it is difficult to achieve a fully spin-polarized current in purely metallic nanocontacts.  An alternative is provided by transition metal oxides, particularly nickel oxide, which, as mentioned above, was predicted to become half metallic in reduced dimensions\cite{jacob2006} and has recently shown evidence of nearly complete spin filtering when realized by oxygenation of nickel break junctions.\cite{vardimon2015}  Several proposals for molecular spin filters have recently been put forward (see e.g.~Refs.~\onlinecite{sanvito2011,zu2013,smogunov2015} and references therein), where spin-filtering behaviour is often connected with magnetoresistance.  The molecular bonding group and its spin-dependent hybridisation with the leads is an important factor determining the performance of molecular spin valves and spin filters.  

\noindent \textbf{References}

\begingroup
\renewcommand{\section}[2]{}

\endgroup

\noindent \textbf{Acknowledgements}

\noindent Work in Trieste partly sponsored by ERC Advanced Grant 320796 -- MODPHYSFRICT.  

\noindent \textbf{Competing financial interests}

\noindent The authors declare no competing financial interests.

\end{text}

\end{document}